# A RADIO SCIENCE EXPERIMENT FOR THE RAMSES MISSION TO APOPHIS

Riccardo Lasagni Manghi[*†], Marco Zannoni[*†], Edoardo Gramigna[*†], Paolo Tortora[*†], Giacomo Paialunga[†], Andrea Negri[‡], Giovanni Cucinella[‡], Pier Luigi De Rubeis[§], Lorenzo Simone[§]

This paper outlines the Radio Science Experiment (RSE) proposed for the RAMSES mission to asteroid (99942) Apophis, which will undergo a close Earth encounter in April 2029. This event provides a unique opportunity to study the asteroid's physical and dynamical changes under strong tidal forces. The experiment leverages a combination of Earth-based radiometric measurements, optical imaging, and inter-satellite links between the RAMSES mothercraft and deployable subcraft in proximity to Apophis. Using high-precision Doppler and optical navigation data, the RSE aims to estimate the asteroid's mass, gravity field, and spin state with unparalleled accuracy, furthering our understanding of near-Earth asteroid evolution and internal structure. Simulation results show the robustness of the proposed mission scenario, highlighting the critical role of multi-probe configurations and novel inter-satellite link technologies in achieving accurate gravity science results.

## INTRODUCTION

On April 13, 2029, the asteroid 99942 Apophis will have a very close encounter with the Earth, transiting the GEO ring. This flyby represents a unique opportunity to observe a well-known potentially hazardous asteroid subject to strong tidal forces. The time-varying orbital and rotational environment can lead to changes in the surface slopes. Depending on the circumstances, this mechanism may drive significant property changes in the asteroid's internal structure and granular motion on its surface. In this context, characterizing the bulk density and its mass distribution within the asteroid nucleus before and after the encounter could represent a critical step towards understanding the evolution history of near-Earth asteroids[1].

In this work, we present the outline of a possible Radio Science Experiment (RSE) onboard the Rapid Apophis Mission for Space Safety (RAMSES) proposed by the European Space Agency (ESA), which is expected to rendezvous with the asteroid in February 2029, right before the Earth close encounter. The objectives of this experiment will include characterizing the overall mass, density, and porosity of the nucleus with an accuracy of less than 1%, determining its spin rate and orientation to less than 1% and 5°, respectively, estimating the extended gravity field and internal structure of the nucleus, and improving its heliocentric trajectory reconstruction.

[*] Centro Interdipartimentale di Ricerca Industriale Aerospaziale (CIRI AERO) Alma Mater Studiorum – Università di Bologna, Via Baldassarre Carnaccini 12, 47121 Forlì, Italy.
[†] Dipartimento di Ingegneria Industriale (DIN), Alma Mater Studiorum - Università di Bologna, Via Fontanelle 40, 47121 Forlì, Italy.
[‡] IMT S.r.l., via Carlo Bartolomeo Piazza, 30, 00161, Rome, Italy.
[§] Thales Alenia Space Italia, via Saccomuro 24, 00131, Rome, Italy

The radio science experiment will combine Earth-based radiometric measurements, namely Doppler, range, and ΔDOR, with optical images collected by the onboard navigation cameras to reach the outlined objectives. Furthermore, the RSE will exploit the radiometric measurements collected through the Inter-Satellite Link with one (or more) subcraft released by the RAMSES mothercraft, providing high-accuracy Doppler measurements at closer orbital distances from the target.

The mission might serve as the first deep space application for a new Inter-Satellite Link Transceiver (ISL-T) for CubeSats, featuring a dedicated, fully coherent Doppler channel. Developed in the framework of the ASI-funded INNOVATOR project[2] and realized by Thales Alenia Space Italia, the new ISL-T is expected to be tested and validated in low-Earth orbit in late 2025 or early 2026, reaching a high level of maturity before the RAMSES integration campaign.

The paper is organized as follows: the first section presents a brief overview of the proposed mission concept and its timeline; the second section introduces the multi-arc covariance analysis, the overall simulation procedure, the dynamical model used for the propagation of the asteroid and spacecraft states, and the measurement model; the third section shows the main results of the analysis, highlighting the contributions of the individual probes to the gravity reconstruction and the role of the optical navigation images in reconstructing the rotational state of Apophis; finally, the last section draws the conclusions and outlines future investigation topics.

**MISSION SCENARIO**

Building on the experience gained with the RSE onboard the Hera mission[3,4,5], this study proposes a concept of operations involving a radio link between RAMSES and one or more deployable subcraft. As shown in Figure 1 and Table 1, the mothercraft is expected to rendezvous with Apophis in late February 2029. During a pre-encounter phase (PRE) of roughly 40 days, the spacecraft will alternate passively safe hyperbolic orbits and hovering boxes at various altitudes between 20 km and 1 km, as shown in Figure 2. During the Earth close encounter phase (CEP), the spacecraft will retreat to a safe distance of roughly 15 km, hovering at a constant Sun phase angle and monitoring the evolution of Apophis with high temporal resolution imaging (up to 1 image per minute). In the post-encounter phase, the spacecraft will again move closer to the asteroid, mirroring the trajectories of the pre-encounter phase.

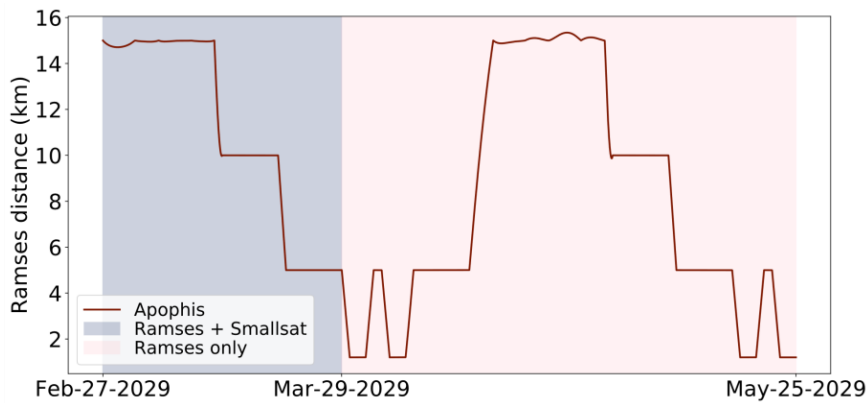

**Figure 1 Ramses-Apophis distance as a function of the mission timeline in the ESA proposed concept of operations.**

The subcraft, released in early March 2029, will transfer to a periodic terminator orbit (PTO) at a distance of roughly 1.25 km from Apophis, where they will operate for one month. Showing stability under significant SRP perturbations, even at low orbital altitudes, these orbits are, in fact, well suited for small-body exploration and high-precision gravity field reconstruction. Furthermore, with a careful selection of the orbital period, PTOs can provide full surface coverage.

After nominal operations, the subcraft will land on the asteroid's surface before the Earth's close encounter. This choice was driven mainly by the following considerations: the dynamical environment near the Earth may affect the stability of the orbit, thus complicating the spacecraft operations significantly; furthermore, some of the scientific payloads to be carried onboard the subcraft may benefit from operating at the surface during the close-encounter. Specifically, proposed payloads include a seismometer and gravimeter, which might help investigate the tidal dynamics, energy dissipation, and surface re-shaping mechanisms during the encounter.

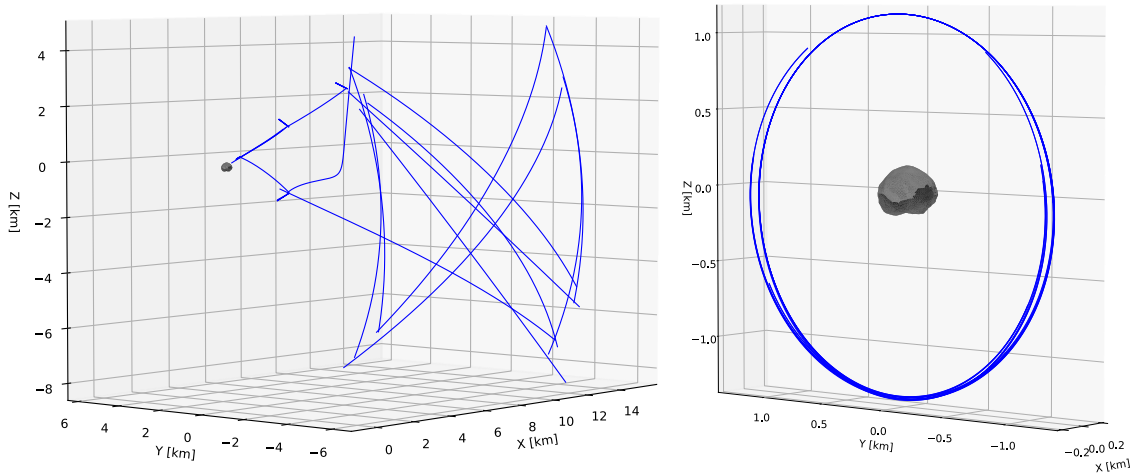

**Figure 2 Orbits of Ramses and subcraft around Apophis. Left Ramses proximity phase; Right: subcraft proximity orbits, corresponding to periodic terminator orbits.**

**Table 1 Reference timeline and simulated orbits for the RAMSES mission. Maincraft: R, subcraft (S)**

| Phase | Interval | Description |
|---|---|---|
| Rendezvous | February 2029 | - |
| Pre-encounter Phase (PRE) | 1 March - 11 April | R: Passively safe hyperbolic arcs (15km); Hovering at (15,10,5,1) km<br>S: PTOs at 1.25 km |
| Close-encounter Phase (CEP) | 12-14 April | R: Hovering at 5 km (Constant Sun phase angle)<br>S: surface operations |
| Post-encounter Phase | 15 April – July | M: hyperbolic + hovering trajectories (TBC)<br>S: surface operations |
| End of operations | August 2029 | - |

## METHODS

A radio science experiment represents a particular application of the orbit determination process in which a series of parameters is estimated that completely define the past trajectory of a spacecraft and enable it to predict its future evolution. For the RAMSES radio science experiment, these parameters include, among others, the mass of Apophis, its extended gravity field expressed in terms of spherical harmonics, its heliocentric trajectory, and its rotational state.

To assess the formal uncertainty of the estimated parameters before the execution of the mission, numerical simulations are run, which follow the same procedure as the actual experiment, using simulated measurements in place of observed data. For this work, a covariance analysis was performed, meaning that the dynamical and observational models used for generating the simulated observables are the same ones used for the estimation process. This approach is often used during mission design phases to understand better how the spacecraft trajectories and the main design parameters affect the experiment's performance[3,6].

However, the real uncertainty associated with the estimated parameters is often larger than the formal value estimated within the simulated orbit determination process due to linearization errors, biases in the estimated values induced by unmodelled physical effects, and correlated measurement errors. A series of conservative assumptions were made to account for these factors and obtain realistic expected uncertainties, including higher noise levels and larger *a priori* uncertainties for the solved-for parameters.

Numerical simulations were performed using NASA/JPL's navigation software, the Mission Analysis, Operations, and Navigation Toolkit Environment (MONTE)[7], whose mathematical formulation is described by Moyer[8], and which is currently used for the operations in NASA's deep-space missions and radio science experiments data analysis[9,10,11,12].

**Dynamical Model**

The gravitational accelerations used for the trajectory propagation of Apophis and of the spacecraft include relativistic point-mass gravity of the Sun, the Solar System planets and their satellites, and the 16 most massive bodies in the main asteroid belt; higher-order gravitational harmonics of the Earth up to degree 30 and of Apophis up to degree 5. Specifically, we used the latest polyhedral shape derived from radar observations[13] to compute Apophis's generalized moments of inertia using the formulation of Hou[14] and assuming a uniform density. The spherical harmonics coefficients were then derived from the inertia tensor using the formulation by Tricarico[15].

The gravitational coefficients and initial state vectors of the Solar System bodies were taken from JPL's DE440 planetary ephemerides. In contrast, the initial state of Apophis was retrieved from JPL's Small Body database[16] in the form of osculating orbital elements.

The rotational model of Apophis was initially described with respect to the ecliptic mean orbit at J2000 (EMO2000) by its pole right ascension $\alpha$ and declination $\delta$ coordinates, which are modeled as linear functions of time. Similarly, the prime meridian $w$ is modeled as a linear function of time starting from an initial zero value at the node and having a constant angular frequency. In addition to this model, which represents the standard for gravity science analysis of planetary targets, a second part of the analysis explores Apophis's non-principal axis (NPA) motion[17] and its gravitational interaction with the Earth's gravity field during the close encounter.

Non-gravitational acceleration due to solar radiation pressure (SRP) was computed using a standard flat-plates model, assuming a Hera-like structure for RAMSES and a 6U CubeSat with deployable solar panel configuration for the subcraft. Furthermore, to simulate unmodelled dynamics from thermal recoil pressure, Earth albedo, and thruster performance errors, we included stochastic accelerations acting in the spacecraft body axes using 18-hour interval batches. The assumed magnitude of stochastic accelerations is particularly relevant to the current concept of operations proposed by ESA for the RAMSES maincraft. During the proximity operations, the spacecraft will undergo several hovering phases, during which the orbit control thrusters will operate almost continuously. While these trajectories are helpful for surface mapping at constant phase angles, the frequent use of thrusters may significantly degrade the accuracy of the spacecraft state reconstruction, thus reducing the potential of gravity science estimation.

Table 2 summarizes the main physical properties of the spacecraft surfaces. SRP modelization errors typically represent one of the primary error sources for non-gravitational accelerations. Local SRP scale factors were estimated for each arc as part of the orbit determination process to mitigate possible modeling errors.

Furthermore, to simulate unmodelled dynamics from thermal recoil pressure, Earth albedo, and thruster performance errors, we included stochastic accelerations acting in the spacecraft body axes using 18-hour interval batches. The assumed magnitude of stochastic accelerations is particularly relevant to the current concept of operations proposed by ESA for the RAMSES maincraft. During the proximity operations, the spacecraft will undergo several hovering phases, during which the orbit control thrusters will operate almost continuously. While these trajectories are helpful for surface mapping at constant phase angles, the frequent use of thrusters may significantly degrade the accuracy of the spacecraft state reconstruction, thus reducing the potential of gravity science estimation.

**Table 2 Shape models and optical coefficients for the RAMSES spacecraft**

| Probe | Component | Area (m$^2$) | Specular | Diffusive |
|---|---|---|---|---|
| Maincraft | High-gain antenna | 2.54 | 0 | 0.327 |
| | Bus (top/side/front) | 3.6 / 3.78 / 4.2 | 0.0735 | 0.2520 |
| | Solar arrays | 4.35 | 0.038 | 0.052 |
| Subcraft | Bus (top/side/front) | 0.0425 / 0.0875 / 0.0278 | 0.0735 | 0.2520 |
| | Solar arrays | 0.1455 | 0.038 | 0.052 |

**Measurement Models**

Simulated observables include two-way Doppler and range measurements in the X-band between the RAMSES and the ESTRACK ground stations, assuming a conservative coverage of 8 h per orbit determination tracking arc, lasting between one and three days. For these measurements, we assumed noise values of 0.07 mm/s for the Doppler at 60 s count time and 43 cm for the range, which includes a safety factor of 2 compared to the typical accuracy for single-frequency links at X-band[18].

Furthermore, two-way range and Doppler measurements at the S-band were simulated for the inter-satellite link between RAMSES and the subcraft. For the ISL measurements, we assumed a 40 % duty cycle, equally spread over the subcraft operational life, corresponding to 2 minutes of tracking every 5 min of operations. Conservative noise values of 0.05 mm/s and 50 cm were used for the Doppler at 60 s count time and for the range, in agreement with preliminary tests on ISL hardware components for the Hera mission[19].

Optical images of the RAMSES navigation camera were acquired outside of tracking periods at a sampling rate of 1 picture every 2 hours, complementing the radiometric data and providing a constraint for the relative orbit reconstruction. The measurements consist of sample and line coordinates of 258 equally spaced landmarks generated on the surface of Apophis. The camera specifications were taken from the Hera AFC camera. A Gaussian noise of 2 pixels was added to each target's sample and line coordinates. For each picture, a camera pointing error is modeled through three successive rotations around the camera axes, estimated in the orbit determination filter starting from an *a priori* uncertainty of $10^{-2}$ deg.

**Filter Setup**

In this study, we adopted a multi-arc approach, commonly used for the orbit determination of missions towards small bodies[20]. This approach combines the data collected during non-contiguous orbital segments or arcs. It analyzes them jointly in a weighted least-square batch filter to produce a single set of solved-for parameters. The solution is represented by the estimated values of the parameters and the corresponding covariance matrix, which provides the formal estimation uncertainty. The solved-for parameters, which are summarized in Table 3, can be broadly divided into *global* parameters, which do not vary during the mission and thus affect all tracking arcs, and *local* parameters, which only affect the measurements of individual arcs.

**Table 3 Filter setup summary**

| Parameter | Type | *A priori* σ | Comments |
| --- | --- | --- | --- |
| *RAMSES (maincraft and subcraft)* | | | |
| Position | Local | 10 km | Spacecraft state with respect to Apophis. Accounts for maneuver errors at the beginning of the arc |
| Velocity | Local | $10^{-4}$ km/s | |
| *Apophis* | | | |
| Position | Global | 100 km | Widely open |
| Velocity | Global | $10^{-5}$ km/s | |
| GM | Global | $2.0 \cdot 10^{-9}$ km$^3$/s$^2$ | 100% of the simulated Apophis mass, derived from the estimated volume from the radar shape model[13] and assuming a bulk density of $1.51 \cdot 10^3$ kg/m$^3$ |
| $J_2$ | Global | $4.1737 \cdot 10^{-1}$ | From the power spectrum of computed values (constant density polyhedron), scaled by a factor of 10. Only the degree and order two are reported for brevity. |
| $C_{21}$ | Global | $1.2191 \cdot 10^{-2}$ | |
| $S_{21}$ | Global | $2.3777 \cdot 10^{-3}$ | |
| $C_{22}$ | Global | $1.4057 \cdot 10^{-1}$ | |
| $\alpha_0$ | Global | 25 deg | Widely open |
| $\delta_0$ | Global | 25 deg | |
| $w_1$ | Global | $5.3325 \cdot 10^{-7}$ deg/s | From rotation period uncertainty, scaled by a factor of 5. |
| *SRP* | | | |
| Scale Factor | Local | 1.0 | Uncertainty is 100 % of the acceleration |
| *Pointing error per picture* | | | |
| 3 Rotations | Local | 10 mdeg | Same as Rosetta |
| *Apophis landmark positions* | | | |
| Radius | Global | - | 10 % of the equivalent radius |
| Lat. / Long. | Global | 5 deg | |
| Scale factor | Global | 0.1 | 10 % size scale, common to all landmarks |

**RESULTS**

**Spacecraft contributions**

A series of simulations were run to assess the radio science experiment performance at the end of the pre-encounter phase as a function of the RAMSES concept of operations. Specifically, the following scenarios were explored, which assess the information content provided by the individual probes and their measurement configuration:

1. *Ramses-only* scenario: only measurements collected by the maincraft are available, namely range and Doppler from the Earth-based tracking link and OPNAV images from the AFC;
2. *+Smallsat* scenario: all measurements collected by the maincraft in the *Ramses-only* scenario are available. Furthermore, range and Doppler ISL measurements are included, using the maincraft-subcraft tracking with one subcraft and assuming a 40% duty cycle.
3. *+2 Smallsat MST* scenario: all measurements collected by the maincraft in the *Ramses-only* scenario are available. Furthermore, range and Doppler ISL measurements are also included, using the maincraft-subcraft tracking (MST) with two subcraft placed on the same PTO at a fixed mean anomaly separation and assuming a 40% duty cycle for each subcraft;
4. *+2 Smallsat SST* scenario: all measurements collected by the maincraft in the *Ramses-only* scenario are available. Furthermore, range and Doppler ISL measurements are included, using the subcraft-subcraft tracking (SST) for a 20% duty cycle and MST for another 20%.

Furthermore, each scenario was simulated using alternative assumptions for the *a priori* uncertainty of the stochastic accelerations acting on the RAMSES maincraft to address the sensitivity of the results to the thruster performance errors. Specifically, we assumed either values of $\sigma_{0,nga} = 10^{-12} km/s^2$ or $\sigma_{0,nga} = 10^{-11} km/s^2$, corresponding to roughly 5% and 50% of the expected SRP magnitude during the proximity phase.

Figure 3 shows the estimated accuracy of Apophis' GM and $J_2$ for the simulated scenarios at the end of the close-encounter phase. It can be seen that all scenarios satisfy the mission requirements of <1% accuracy for the mass estimation, with values ranging from 0.2% in the *Ramses-only* scenario to 0.04% in the *+2 Smallsat SST* scenario.

Conversely, we notice that the $J_2$ is not observable in the *Ramses-only* scenario when the initial uncertainty of the stochastic accelerations is high. Adding a single deployable subcraft with ISL capabilities improves accuracy to roughly 10%. Adding a second subcraft in the SST configuration further improves the estimation to 2% accuracy.

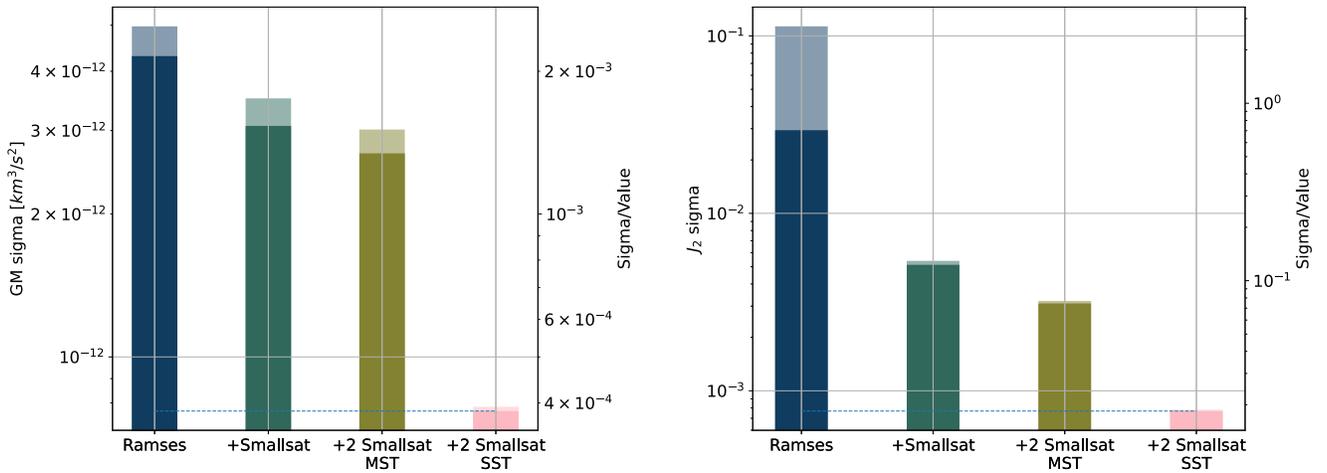

**Figure 3 Estimated accuracy (1σ) of Apophis GM (left) and $J_2$. (right). Transparent bars: scenario with $\sigma_{0,nga} = 10^{-11} km/s^2$; full bars: scenario with $\sigma_{0,nga} = 10^{-12} km/s^2$.**

Similar considerations can be drawn from Figure 4, showing Apophis' gravity spectra for a spherical harmonic expansion up to degree and order 5. Specifically, the continuous black line represents the root mean square (RMS) value of the Stokes coefficients of all orders for a given degree expansion. Similarly, the colored lines represent the RMS value of the estimated uncertainty of the Stokes coefficients at the end of the close-encounter phase. In the *Ramses-only* scenario, degree-two harmonics can be marginally estimated (~70% accuracy) only when a good knowledge of the spacecraft NGA is available. Adding one or more subcraft produces an order of magnitude improvement in the degree-two estimation and allows for observing the degree-three harmonics, thanks to the lower altitude of the PTOs and their different observing geometry, which helps remove the ambiguity in the gravity field reconstruction around the poles.

We can also notice that the dependence of gravity reconstruction on the NGA knowledge is strongly reduced in the presence of one or more subcrafts, particularly when using the subcraft-subcraft ISL configuration. The *+2 Smallsat SST* scenario is, therefore, more robust to variations in the RAMSES concept of operations and is fully compatible with the current orbit design shown in Figure 2 that relies on hovering for the Apophis surface mapping. In the following section, we will thus use this latter as the reference scenario for the RAMSES mission.

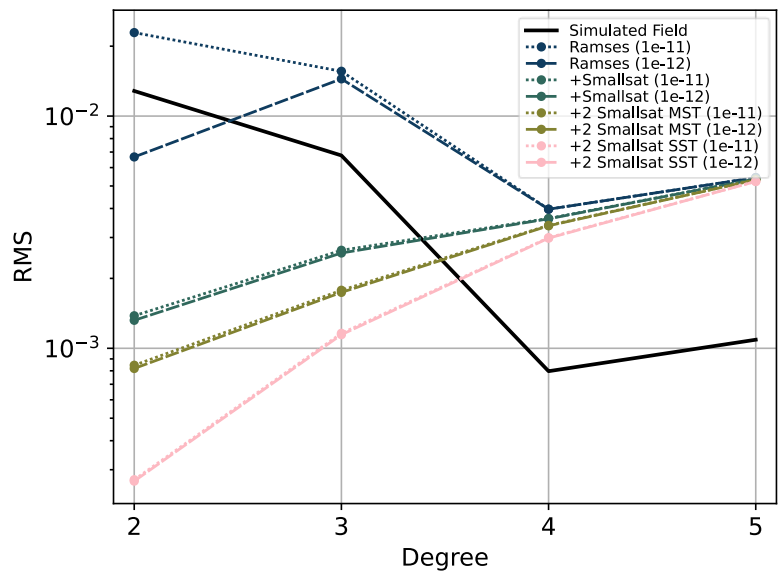

**Figure 4 Apophis gravity spectra. The black continuous line is the power spectrum of the simulated field (RMS of the Stokes coefficients of all orders for a given degree of spherical harmonics expansion). Colored lines: estimated formal uncertainties (1σ) at the end of the close-encounter phase. Dashed lines: $\sigma_{0,nga} = 10^{-12} km/s^2$; dotted lines: $\sigma_{0,nga} = 10^{-11} km/s^2$.**

**Apophis NPA spin state**

Previous studies on the spin state of Apophis employing ground-based lightcurves[17] and radar observations[13] indicate that the asteroid is likely characterized by an NPA rotational state. While this feature may complicate the operations related to shape reconstruction and surface mapping, it represents an opportunity for improving gravity science since results. In a torque-free attitude motion, characteristic frequencies of the angular rates around the principal axes of an extended body are tied to its moments of inertia and, consequently, to its internal density distribution. Furthermore, tidal torques from the Earth's gravity field will significantly modify the spin state of Apophis, with an overall effect that is expected to depend on the local attitude state at the closest approach[21,22]. Characterizing the spin state of Apophis throughout the RAMSES mission is, therefore, of paramount importance.

In this second part of the analysis, we modified the dynamical setup by propagating Apophis' rotational state using Euler's equations of motion for rigid bodies and starting from the initial excited conditions derived by Brozović et al.[13], which are summarized in Table 4. The initial attitude state at the reference epoch is given below using the convention defined by Kaasalainen[23]. The state was propagated forward in time to get the initial conditions at RAMSES arrival. The Earth's gravity field induced external torque was modeled using the gravity gradient and a degree 30 spherical harmonics expansion.

**Table 4 Apophis spin state at the reference epoch Dec. 23, 2012, 14:14 UTC**

| Parameters | *A priori* value | Unit | Description |
| --- | --- | --- | --- |
| $\omega_l, \omega_i, \omega_s$ | [96.506, 50.799, 264.953] | deg/day | Angular rates of the Apophis long, intermediate, and short body axes |
| $\alpha_L, \delta_L$ | [250, -75] | deg | Right ascension and declination of the angular momentum vector in the J2000 ecliptic reference frame |
| $\phi_0, \theta_0, \psi_0$ | [133.8, 17.8, 55.6] | deg | 3-1-3 Euler angles of Apophis body frame with respect to the angular momentum vector at the reference epoch |

Figure 5 shows the gravity spectra obtained with the new dynamical model that includes Apophis' attitude integration. Specifically, we consider the Ramses only and +2Smallsat SST scenarios here, representing the most extreme CONOPS cases. Both scenarios assume an *a priori* uncertainty of $\sigma_{0,nga} = 10^{-11} km/s^2$ for the RAMSES non-gravitational accelerations. Furthermore, we assess the sensitivity of the results to the sampling frequency of the optical images from the navigation camera onboard RAMSES.

By comparing *Ramses only* curve with the corresponding one in Figure 4, we notice that the NPA state of Apophis and its interactions with the Earth's gravity field allow us to improve the degree-two gravity reconstruction by roughly an order of magnitude by enabling the estimation of the body's moments of inertia. Even though the direct estimate of the higher-order harmonics is not possible through the rotational state reconstruction, increasing the optical image frequency allows us to reduce the spacecraft's state uncertainty, thus reducing the uncertainty of degree three. Similarly, a factor 2 reduction is observed in the degree-two harmonics for the *+Smallsat SST* case with respect to the standard results in Figure 4. However, this scenario seems less affected by the frequency of the RAMSES navigation images since most of the gravity information content is provided directly by the subcraft.

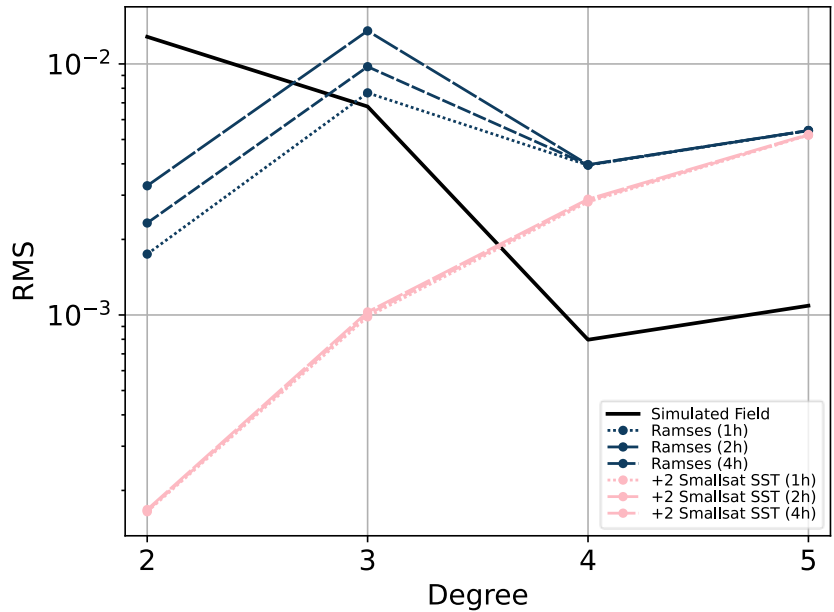

**Figure 5 Apophis gravity spectra.** Black continuous line: power spectrum of the simulated field. Colored lines: estimated formal uncertainties (1σ) at the end of the close-encounter phase for the *Ramses only* (blue) and *+2Smallsat SST* (pink) scenarios. Markers indicate the frequency of optical navigation imaged: long-dashed (4 h), dashed (2h), dotted (1h). All cases assume a RAMSES *a priori* uncertainty of the NGAs of $\sigma_{0,nga} = 10^{-11} km/s^2$.

Finally, Figure 6 shows the estimated accuracy of the spin state reconstruction as a function of time for the nominal scenario (*+2 Smallsat SST*). We observe that the absolute accuracy for all three axes is always below 1° during and after the Earth C/A event, corresponding to less than 3 m displacements on the surface of an equivalent-volume sphere. Furthermore, the rotational speed can be estimated with an accuracy of 1 deg/h for the short and intermediate axes and 3 deg/hour for the long axis, which satisfies the 1% relative accuracy requirement.

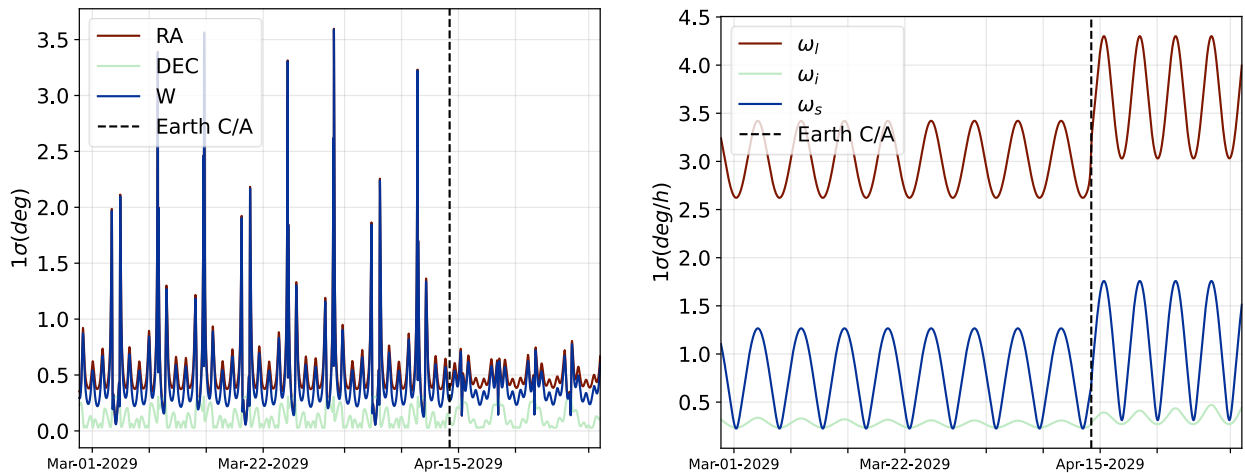

**Figure 6 Estimated accuracy (1σ) of Apophis' body fixed frame in the form of 1-2-3 Euler axes (left) and angular rates (right). Vertical dashed line: epoch of the Earth's closest approach.**

## CONCLUSIONS

This work presents an outline of the proposed Radio Science Experiment for the ESA's RAMSES mission to asteroid (99942) Apophis and shows its expected results in terms of gravity estimation at the end of the Earth close-encounter phase for different concepts of operation. Specifically, we compare a *Ramses-only* scenario, in which the probe relies only on ground-based tracking and optical images for navigation, with scenarios including one or two deployable subcraft placed at 1.25 km terminator orbits with inter-satellite link capabilities.

The covariance analysis results indicate that the *Ramses-only scenario satisfies the mass estimation requirement with a relative accuracy of 0.2%, which improves up to 0.04% in the* presence of two deployable subcraft. However, in the *Ramses-only* scenario, the degree-two harmonics estimation highly depends on the knowledge of the probe's non-gravitational accelerations, which might be severely degraded during the currently planned hovering phases. Adding one or two deployable subcraft enables the gravity field estimation up to degree and order 3, improving the overall science return and increasing the robustness of the mission.

Orbit determination simulations were also performed by propagating the asteroid's body frame from its estimated NPA spin state and including the tidal torques induced by the Earth's gravity field during the 2029 close encounter. Preliminary results indicate that Apophis' NPA spin state might improve the gravity science reconstruction by enabling the estimation of the asteroid's moments of inertia, which are strictly coupled to the characteristic frequencies of precession and nutation around the angular momentum direction. Furthermore, the proposed concept of operations satisfies the mission requirements in terms of spin state reconstruction, which is paramount for characterizing tidal reshaping and surface motion in the aftermath of the Earth's flyby.

A complete characterization of the expected results would require performing a sensitivity analysis to the actual spin state at the time of the Earth's closest approach, whose value has been shown to impact the NPA state after the encounter significantly. However, this analysis is beyond the scope of the current work and will be the subject of future investigations.

Further improvements in the orbit determination accuracy may be obtained by exploiting the optical images collected by the subcraft's onboard cameras, which were not considered for this analysis, or extending the subcraft's ISL operations after the surface landing. The ISL Doppler data at the surface may, in fact, improve the Apophis attitude reconstruction significantly, with clear benefits for the local acceleration estimation and inertia moments reconstruction.

Future work may also focus on optimizing the concept of operations for the subcraft by employing alternative orbits, such as Resonant Terminator Orbits (RTOs) defined in the augmented Hill three-body problem, which may potentially improve the surface coverage and provide different geometric conditions for optical navigation images[24].

## ACKNOWLEDGEMENTS

EG, RLM, MZ, and PT wish to acknowledge *Caltech* and the *NASA Jet Propulsion Laboratory* for granting the University of Bologna a license to an executable version of MONTE Project Edition S/W.